# Analysis of photon characteristics in anticorrelation of a Hong-Ou-Mandel dip for on-demand quantum correlation control

Byoung S. Ham

Center for Photon Information Processing, School of Electrical Engineering and Computer Science, Gwangju Institute of Science and Technology

123 Chumdangwagi-ro, Buk-gu, Gwangju 61005, S. Korea

(Submitted on May 25, 2021; bham@gist.ac.kr)

**Abstract:** Over the last several decades, quantum entanglement has been intensively studied for potential applications in quantum information science. The Hong-Ou-Mandel (HOM) dip is the most important test tool for direct proof of entanglement between paired photons, whose coincidence detection results in anticorrelation due to photon bunching on a beam splitter. Although anticorrelation is due to destructive quantum interference between paired photons, a wavelength-sensitive interference fringe has never been observed in any HOM-type experiments. Here, a typical HOM dip is investigated for entangled photon pairs generated by parametric down conversion processes (SPDC) to understand fundamental physics of anticorrelation. In addition, a pure coherence optics-based Hong-Ou-Mandel scheme is proposed and analyzed for general understanding of anticorrelation in an interferometric system. This study sheds light on deterministic quantum correlation control and opens the door to potential applications of on-demand quantum information science.

**Introduction**

Since the seminal papers on the Bell inequality [1] and anticorrelation, the so-called Hong-Ou-Mandel (HOM) dip [2], quantum entanglement [3-5] has been intensively studied for potential applications on quantum computing [6-10], quantum cryptography [11-15], and quantum sensing [16-20]. The anticorrelation of the HOM dip results from two-mode photon bunching on a beam splitter (BS), where the zero coincidence-detection rate is direct proof of the nonclassical features of two-mode entanglement [1-5]. A HOM dip has also been observed from two independent and incoherent light sources, e.g., the sunlight and quantum dots. [21,22]. Although photon bunching in a HOM dip is because of destructive quantum interference between the paired photons, a typical interference fringe has never been observed in any HOM-type experiment. Thus, the quantum correlation of a HOM dip has become a weird phenomenon, resulting in nondeterministic measurement-based physics based on particle characteristics. Here, a specific phase relation between the paired photons used for a typical HOM dip is investigated to find the fundamental mechanism of anticorrelation and what makes the interference fringe disappear. Moreover, a coherent light-based HOM-type scheme is proposed and analyzed for a general understanding of anticorrelation on a BS. Finally, a deterministic quantum correlation technique is discussed for potential applications of on-demand quantum information science.

Based on Heisenberg's uncertainty principle for a photon with paired conjugate variables, e.g., time and frequency, $g^{(2)}$ intensity correlation is closely related with $g^{(1)}$ amplitude correlation according to general coherence optics: $g^{(2)} = g^{(1)} + 1$ [23]. Thus, anticorrelation of $g^{(2)}(\tau = 0) = 0$ can be fully explained by coherence optics with $-1 < g^{(1)} < 1$. In that sense, the wave nature of a photon is well suited to interpret the HOM-type anticorrelation based on destructive quantum interference between two paired photons without violating quantum mechanics originated in the wave-particle duality. According to the Copenhagen interpretation, both features of the particle and wave natures of a photon are mutually exclusive. Based on intensive studies for the Born rule test over the last decade [24-28], both exclusive natures of a single photon result in no difference in an interferometric system, supporting the Copenhagen interpretation. Thus, the coherent approach for anticorrelation of a HOM dip can provide a comprehensive understanding on the quantum feature. Compared with the wave approach, the conventional particle approach results in nondeterministic quantum correlation such as in probabilistic four-different types of Bell states between two-mode entangled photons [29]. This results in an extreme inefficiency in quantum information processing for a multi-mode entangled system, which is necessary for many-node quantum internet [30,31]. Although a single photon's phase is nondeterministic according to the uncertainty principle, a relative phase between two photons can be definite without violating quantum mechanics as demonstrated in anticorrelation [32,33] as well as in the Franson-type Bell inequality [34]. This is the physical origin as to how the wave approach is powerful to understand quantum features in the present investigation.

**Results**

Figure 1 shows a general scheme of coincidence measurements in a HOM-type quantum experiment, where the photon characteristics of the input photons, $E_s$ and $E_i$, determine the detailed physics of anticorrelation. For this, we take pure wave nature of photons governed by coherence optics [23]. Based on two distinct analyses of photon characteristics for the same anticorrelation, we derive two necessary requirements for quantum entanglement observed in HOM-type experiments. For this, independently coherent optical systems are additionally introduced and analyzed for the same anticorrelation. Thus, a deterministic method of quantum feature generation is introduced and discussed for future on-demand quantum information processing. As a result, conventional vagueness or weirdness on quantum correlation is ruled out.

*2-1. A classical approach with independently coherent light sources*

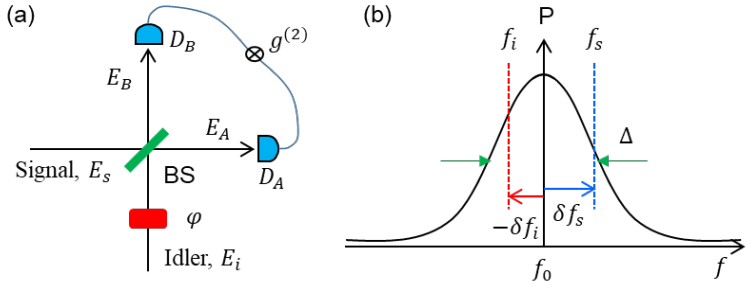

**Fig. 1.** Schematic of anticorrelation on a BS. (a) original scheme. (b) Spectral distribution of independent coherent photons. BS: nonpolarizing 50/50 beam splitter. $D_k$: single photon detector for $E_k$. $f_0$: center frequency of the Gaussian distributed photons. φ: a path-length dependent phase shift.

For the photon characteristics in a HOM-type experiment, a classical approach with independently coherent light sources is firstly investigated. Here, all photons are described as waves of coherence optics governed by Maxwell's equations. To satisfy independency between two input fields (photons) $E_s$ and $E_i$ in Fig. 1(a), two independent but identical lasers are considered. Thus, photons from each laser acting as either signal or idler for the beam splitter (BS) in Fig. 1(a) are coherent. However, photons between two lasers are incoherent in general due to mostly cavity alignment, resulting in slightly different center wavelengths. As shown in Fig. 1(b), each photon's frequency is random within the spectral bandwidth ∆ of each laser. With optical attenuation for each laser, nearly sub-Poisson distributed photons can be easily achieved (see the Supplementary Information), where the spectral bandwidth ∆ also applies to each photon's bandwidth as experimentally demonstrated recently [35]. This same coherence feature between a quantum particle and classical light is not new but strongly supported by Born rule regarding self-interference [24-28]. The coincidence detection between two output fields of the BS is limited by the input photons' characteristics such as coherence length $l_c$: $l_c \tau_c = c$, where $l_c$, $\tau_c$, and c are coherence length, coherence time, and the speed of light, respectively.

Using matrix representations of coherence optics, the following output fields are obtained for Fig. 1:

$$\begin{bmatrix} E_A^j \\ E_B^j \end{bmatrix} = \frac{1}{\sqrt{2}} \begin{bmatrix} 1 & i \\ i & 1 \end{bmatrix} \begin{bmatrix} 1 & 0 \\ 0 & e^{i\varphi} \end{bmatrix} \begin{bmatrix} E_s^j \\ E_i^j \end{bmatrix}$$

$$= \frac{E_s^j}{\sqrt{2}} \begin{bmatrix} 1 & ie^{i\varphi} \\ i & e^{i\varphi} \end{bmatrix} \begin{bmatrix} 1 \\ e^{i(\delta_j + \zeta_j)} \end{bmatrix}$$

$$= \frac{E_s^j}{\sqrt{2}} \begin{bmatrix} 1 + ie^{i(\varphi + \delta_j + \zeta_j)} \\ i + e^{i(\varphi + \delta_j + \zeta_j)} \end{bmatrix}, \qquad (1)$$

where $E_s^j(r,t) = E_0 e^{i(f_s^j \tau - k_0^j r + \delta_j + \zeta_j)}$. The $\zeta_j$ stands for the initial phase difference, while $\delta_j$ is the frequency detuning−caused phase difference between the j$^{th}$ signal and idler photon pair: $\delta f_{is} = \delta f_i - \delta f_s$. In equation (1), the phase control of $\varphi(\tau)$ for $E_i^j$ applies only for $\delta_j$, where $\delta_j(\tau)$ $(= \delta f_{is}\tau)$ causes a different value for each j$^{th}$ photon pair. Due to cavity optics, the spectral variation of $\zeta_j$ cannot be greater than the spectral bandwidth $\Delta$ but also be negligibly small to satisfy the coherent light feature of the laser. As a result, the following output intensities are obtained with Gaussian photon distribution, $G(\delta f_j)$:

$$I_A^j(\tau, \delta f_j) = I_0[1 - sin(\varphi(\tau) + \zeta_j - \delta f_{is}\tau)G(\delta f_j)], \tag{2}$$

$$I_B^j(\tau, \delta f_j) = I_0[1 + sin(\varphi(\tau) + \zeta_j - \delta f_{is}\tau)G(\delta f_j)], \tag{3}$$

where $I_0 = E_s E_s^*$. For the coincidence detection rate at $\tau = 0$, each output intensity rate (average or mean over the spectral bandwidth) becomes equal regardless $\delta f_{is}$. For $\tau \neq 0$, each intensity rate is dependent upon $\delta f_{is}\tau$ and behaves oppositely each other if $\delta f_{is}\tau < \pi$. For $\delta f_{is}\tau \sim \pi$, the rates of equation (2) and (3) reach the same value at $\langle I_0 \rangle$. Usually for independent two identical lasers, however, the condition of $\zeta_j \ll \pi$ is not generally accepted due to slight different cavity alignments. Even in this case, intensity rates of equations (2) and (3) are $\delta f_{is}\tau$ dependent showing a modified broader fringe. Thus, the average intensities of equations (2) and (3) behave oppositely.

The corresponding intensity correlation or coincidence detection for equations (2) and (3) averaged all over the spectrally distribution of photon pairs is as follows:

$$\langle R_{AB}(\tau) \rangle = \langle I_A^j(\tau) I_B^j(\tau) \rangle = \langle I_0^2 \rangle \sum_j^N \langle cos^2(\varphi(\tau) + \zeta_j - \delta f_j t) G^2(\delta f_j) \rangle. \tag{4}$$

Due to the $\tau$ −dependent relative phase $\delta f_{is}\tau$, equation (4) is the same as the intensity rate product of equations (2) and (3). For $\tau = 0$, $\langle R_{AB}(\tau) \rangle = 1$ violates anticorrelation but satisfies classical intensity correlation. As $\tau$ increases, the effect of spectral distribution of $\delta f_{is}$ results in gradual decrease until it reaches at $\langle R_{AB}(\tau) \rangle = \langle I_0^2 \rangle/2$. This is the classical lower bound in intensity correlation: $g^{(2)}(\tau \gg 1) = 1/2$: $g^{(2)}(\tau) = \frac{\langle I_A(t) I_B(t+\tau) \rangle}{\langle I_A(t) \rangle \langle I_B(t+\tau) \rangle}$. If $\zeta_j \geq \pi$ is considered between two input lasers, then incoherence optics dominates regardless of $\delta f_j t$ for both mean intensities: $\langle I_A^j(\tau, \delta f_j) \rangle = \langle I_A^j(\tau, \delta f_j) \rangle = \langle I_0 \rangle$ and $\langle R_{AB}(\tau) \rangle = \langle I_0^2 \rangle/2$.

The condition of $\zeta_j \ll \pi$ among photons from the same laser is always satisfied by the cavity optics, otherwise there is no lasing. If the initial phase different $\zeta_j$ can also be made small enough between two lasers fixed for equation (4), the anticorrelation condition of $\langle R_{AB}(\tau) \rangle = 0$ can be achieved. Such an ensemble coherence control has already been experimentally demonstrated using a set of etalons, where the etalon plays a role of cavity optics, resulting in coherence between two light sources [19]. This is the unspoken secretes in the anticorrelation observations with independent light sources, where the mutual coherence is controlled by an etalon-based cavity optics or well spectrally filtered system.

*2-2. A quantum approach with entangled photon pairs*

For the quantum approach, a typical HOM dip case is analyzed for the SPDC-generated entangled photon pairs. According to the $\chi^{(2)}$ nonlinear optical process of SPDC, both laws of energy conservation and phase matching must be satisfied, resulting in strong phase correlation among three of pump, signal, and idler photons [2,36]. From the energy conservation law, the following detuning relationship is automatically satisfied for the j$^{th}$ photon pair: $f_s^j = f_0 + \delta f_j$ and $f_i^j = f_0 - \delta f_j$. Thus, all SPDC-generated entangled photons are symmetrically detuned by $\pm \delta f_j$ ($\delta f_i = -\delta f_s$) across the half pump frequency $f_0$ (see Fig. 1(b)). From the $\chi^{(2)}$ −based strong phase correlation, the paired entangled photons should satisfy a certain phase relation between them with respect to the pump photon. Due to spontaneous emission process, however, the detuning $\delta f_j$ in each photon pair is random within the bandwidth $\Delta$ of the Gaussian distribution. The initial phase difference ($\zeta_j$) may be dependent upon the pump bandwidth which is much narrower than $\Delta$.

Like the independent coherent photon case in equation (1), the following output fields are obtained for the j$^{th}$ entangled photon pair:

$$\begin{bmatrix} E_A^j \\ E_B^j \end{bmatrix} = \frac{1}{\sqrt{2}} \begin{bmatrix} 1 & i \\ i & 1 \end{bmatrix} \begin{bmatrix} 1 & 0 \\ 0 & e^{i\varphi'} \end{bmatrix} \begin{bmatrix} E_s^j \\ E_i^j \end{bmatrix}$$

$$= \frac{E_s^j}{\sqrt{2}} \begin{bmatrix} 1 & ie^{i\varphi'} \\ i & e^{i\varphi'} \end{bmatrix} \begin{bmatrix} 1 \\ e^{i(\zeta_j - 2\delta_j)} \end{bmatrix}$$

$$= \frac{E_s^j}{\sqrt{2}} \begin{bmatrix} 1 + ie^{i(\varphi' + \zeta_j - 2\delta_j)} \\ i + e^{i(\varphi' + \zeta_j - 2\delta_j)} \end{bmatrix}, \tag{5}$$

where $\varphi'$ represents the unknown and intrinsic phase difference between the signal and idler photons in SPDC. As a result, the following output intensities are finally obtained:

$$I_A^j(\tau) = I_s^j [1 - \sin(\varphi' + \zeta_j - 2\delta f_j \tau) G(\delta f_j)], \tag{6}$$

$$I_B^j(\tau) = I_s^j [1 + \sin(\varphi' + \zeta_j - 2\delta f_j \tau) G(\delta f_j)], \tag{7}$$

where $\zeta_j$ and $\varphi'$ are independent of $\tau$. For a fixed $\varphi'$ value and $\zeta_j \ll \pi$, equations (6) and (7) vary only by $\delta_j$ ($= \delta f_j \tau$). Thus, equations (6) and (7) are basically the same as equations (2) and (3) except for the symmetric detuning between the paired photons. No wavelength sensitive interference fringe exists in the mean output fields' intensity correlation due to $\delta f_j \tau$ effect (analyzed in Figs. 2 and 3).

From equations (6) and (7), the resultant intensity correlation (coincidence detection) rate $\langle R_{AB} \rangle$ is as follows:

$$\langle R_{AB}(\tau) \rangle = \langle I_s \rangle \langle \cos^2(\varphi' + \zeta_j - 2\delta f_j \tau) G(\delta f_j) \rangle, \tag{8}$$

where $\zeta_j$ effect may be negligibly small, otherwise anticorrelation degrades (analyzed in Fig. 4).

For the observed results of $\langle R_{AB}(\tau = 0) \rangle \sim 0$ in an ideal HOM experiment, the intrinsic phase difference $\varphi'$ between paired photons must be $\varphi' = \pm \pi/2$. This also confirms that $\zeta_j \sim 0$. Thus, the intrinsic phase difference $\pm \pi/2$ between entangled photons generated from SPDC is successfully derived as the first solution for the bipartite entanglement condition on a BS for anticorrelation. Depending on the sign in $\varphi' = \pm \pi/2$, however, the output intensities are swapped according to equations (6) and (7), even though it does not affect equation (8). This is the origin of detection randomness in each port of the BS in Fig. 1(a), resulting in $\langle I_A(\tau) \rangle = \langle I_B(\tau) \rangle = \langle I_0 \rangle$. This is the second solution of the nonclassical feature of a HOM dip based on SPDC-generated entangled photon pairs (discussed later).

Regarding the $\pm \pi/2$ phase difference between the signal and idler photons generated from SPDC process, its origin can also be explained in a three-level atomic system composed of ground, intermediate, and excited states using density matrix equations via a Rabi cycle. In this closed-cycle of an optical excitation, the excited atom experiences a $\pi$−phase shift with respect to the ground one. By the decay process, the returned atom to the ground state must gain another $\pi$, resulting in no phase change for a complete Rabi cycle. Thus, an intermediate state atom results a $\frac{\pi}{2}$−phase shift between the generated signal and idler photons as derivd in equation (8). In an entangled trap-ion pair, the $\frac{\pi}{2}$−phase shift between generated paired photons has also been observed [33].

*2-3. Numerical calculations for anticorrelation*

From equation (8) for $\varphi' = \pm \pi/2$ and $\zeta_j = 0$, the coincidence detection rate $\langle R_{AB}(\tau) \rangle$ for the SPDC-generated photon pairs is numerically calculated and analyzed in Fig. 2 for a typical HOM-type experiment. Figure 2(a) shows Gaussian distributed entangled photon distribution. Figure 2(b) is the resulting coincidence detection rate, $R_{AB}(\tau, \delta f_j)$, where $\delta f_j$ is the frequency detuning of the j$^{th}$ photon-pair from the center frequency of the distribution in Fig. 2(a). The delay time $\tau$ is for the idler as shown in Fig. 1(a). Figures 2(c) and 2(d) are for individual output intensities for Fig. 2(b), where the direction randomness of the signal and idler photons is not included intentionally for a while, otherwise they are uniform as analyzed above regardless of the photon detuning (discussed in Fig. 3). Thus, Fig. 2 also represents equation (4) of the modified classical case for $\zeta_j = 0$. Figure 2(e) is for mean output intensities by averaging for all over the spectral detuning $\delta f_j$ in Figs. 2(c) and 2(d), respectively. On the contrary, Fig. 2(f) is for a reduced (filtered) spectral range from $\pm 2\Delta$ to $\pm \Delta$ across the center of Fig. 2(a). The sidebands' wiggles have already been experimentally observed in the coincidence measurements for the spectrally reduced case [37-39], supporting for the justification and validity of the present analyses.

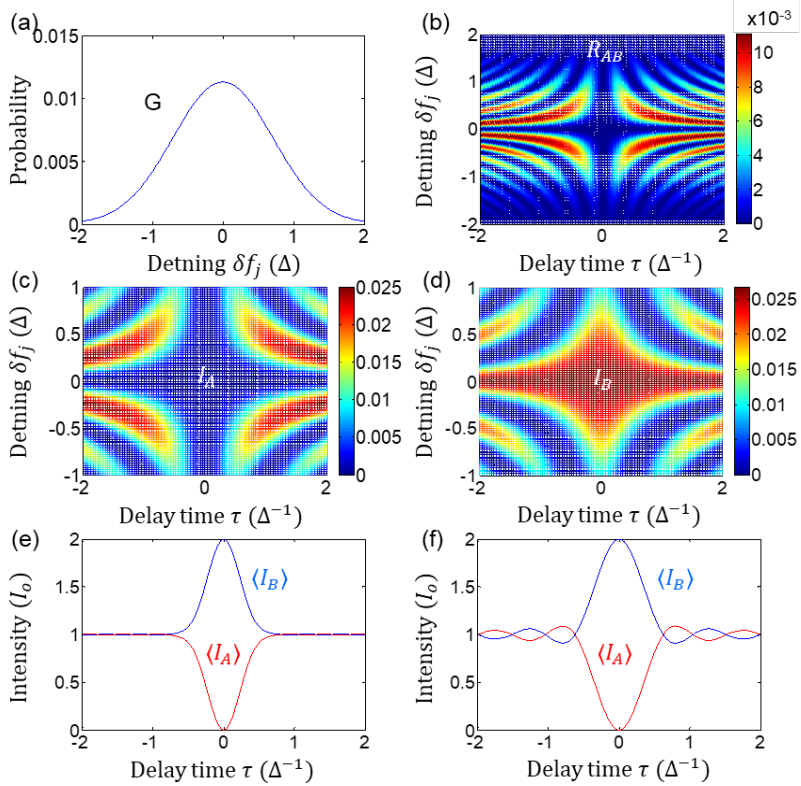

Fig. 2. Numerical calculations for Fig. 1 with SPDC-generated entangled photon pairs without randomness of output directions. (a) Gaussian distribution. b) Coincidence detection $R_{AB}(\tau)$. (c) Intensity $I_A$. (d) Intensity $I_A$. (e) Mean intensity for $4\Delta$ of (a). (f) Mean intensity for $2\Delta$. (g) Second-order intensity correlation $g^{(2)}(\tau)$ for (e). (h) $g^{(2)}(\tau)$ with randomness.

Figure 3 shows numerical calculations for the relationship between coincidence detection $R_{AB}(\tau, \delta f_j)$ and intensity correlation $g^{(2)}(\tau)$ for Figs. 2(b) and 2(e). As expected from equations (4) and (8), nonclassical feature of zero coincidence detection is achieved with a proper condition of $\varphi'$ ($\varphi$) as shown in the dotted green curve in Fig. 3(a). However, the oppositely behaved features of the output intensities in Fig. 2(e) prohibit the nonclassical definition of $g^{(2)}(\tau) < 0.5$ as shown in the blue curve in Fig. 3(a). In equation (8), however, the oppositely detuned frequencies of the signal and idler photons from SPDC can be swapped due to the random detuning by definition. Thus, the output intensities are also swapped randomly, where the photon bunching direction becomes random, too, resulting in $\langle I_A(\tau) \rangle = \langle I_B(\tau) \rangle = \langle I_0 \rangle$. Based on this condition, Fig. 3(b) is for each SPDC-generated photon's output intensities. Figure 3(c) shows comparison between Fig. 2 (red and green dotted curves) and Fig. 3(b) (blue curve) at $\tau = \Delta^{-1}$. The blue curve is the same as the average of both dotted curves resulting from the randomness of entangled photon detections. This detection randomness can also be achieved by classical manipulations of random switching of input fields [40]. This is the powerful benefit of the coherence approach for coherence control of quantum features (discussed in Discussion).

Figure 3(d) is for $g^{(2)}(\tau)$ calculations for Fig. 2(b) under the condition of Fig. 3(b), where the coincidence detection rate is independent of the bunched photon's output direction control in Fig. 1(a). Thus, the nonclassical feature of $g^{(2)}(\tau) < 0.5$ (blue curve) is also satisfied, where the green dotted curve $(R_{AB}(\tau))$ exactly coincides with $g^{(2)}(\tau)$. Thus, the present analysis of anticorrelation for entangled photon pairs from SPDC is successfully demonstrated for a typical HOM dip with numerical supports. From this, two major requirements for the entangled photon pairs have been sought analytically and numerically. From the analyses in Figs. 2 and 3, the nonclassical feature of a HOM dip is not mysterious anymore but deterministic by the phase relation governed by the wave nature of photons without violating quantum mechanics. The uniform intensity in each output field is a necessary condition not for $R_{AB}(\tau)$ but for $g^{(2)}(\tau) < 0.5$. In any way, the coincidence detection rate $\langle R_{AB}(\tau)\rangle$ with a nonclassical feature is independent of $g^{(2)}(\tau)$ as shown in Figs. 3(a) and 3(d). The definition of entanglement may be discussed elsewhere.

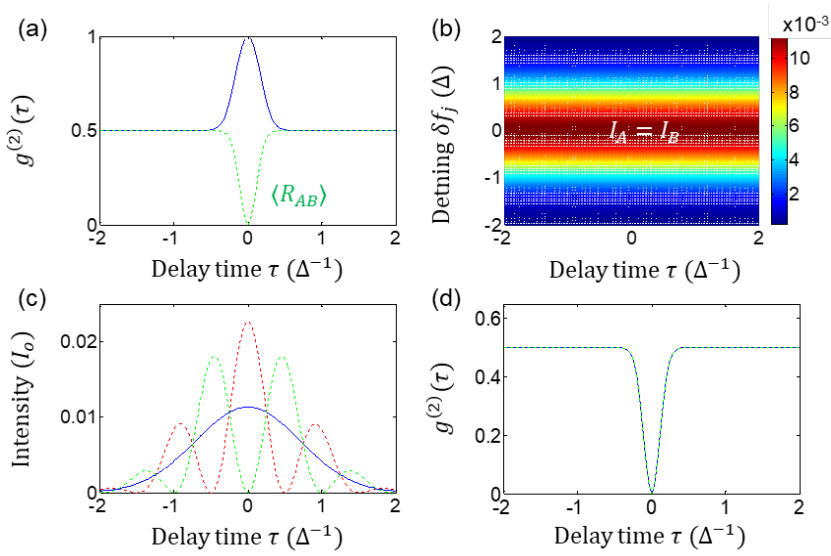

Fig. 3. Numerical simulations for the second-order intensity correlation $g^{(2)}(\tau)$. (a) For Figs. 2(b) and 2(e). (b) For the original SPDC with directional randomness. (c) Green-dotted (Fig. 2(c)), Red-dotted (Fig. 2(d)), Blue (Fig. 3(b)) at $\tau = \Delta^{-1}$. (d) $g^{(2)}(\tau)$ for the original SPDC with directional randomness. Green dotted: $\langle R_{AB}\rangle$.

Figure 4 shows dephasing effect on the coincidence detection for Fig. 1, where the initial phase difference parameter $\zeta_j$ is varied from $\zeta_j = 0$ in Figs. 2 and 3 to $\zeta_j = \pi$ (see the Supplementary Information). For this increasing $\zeta_j$ in equations (4) and (8), $\langle R_{AB}(0)\rangle$ results in gradual decoherence (see the left panel). At $\zeta_j = \pi/2$, $\langle R_{AB}(0)\rangle$ reaches already at complete incoherence of the classical lower bound. The right panel of Fig. 4 shows each $R_{AB}(\tau, \delta f_j)$ for $\zeta_j = \pi/4$ of the left panel. The loss of nonclassicality in two-photon correlation is due to the phase decoherence between the paired photons.

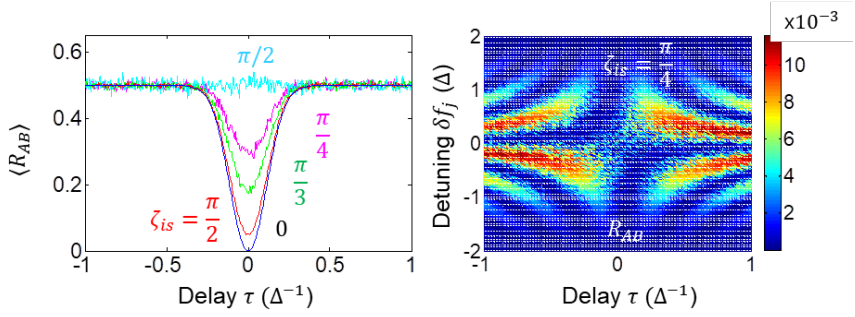

Fig. 4. Numerical calculations for equation (4). (left) $\langle R_{AB}\rangle$ for different $\zeta_{is}$s. (right) $R_{AB}$ for $\zeta_{is} = \pi/4$.

**Discussion**

We analyzed anticorrelation on a BS using different aspects of light sources to understand the fundamental physics of nonclassicality, where a classical case of independently coherent photon pairs from two identical lasers was compared with a quantum case of SPDC-generated entangled photon pairs. In the coincidence detection rate between two output fields from the BS, a smooth and broad single anticorrelation dip, the so-called HOM dip, was analyzed and numerically demonstrated to find definite solutions of anticorrelation using pure coherence optics of the wave nature of a photon. Depending on the coherent spectral filtering of the input photons for the classical case, the observed fringes in both side bands of the dip also clearly demonstrated the physical origin of this quantum feature. The HOM type anticorrelation was clearly analyzed for a specific phase relation between entangled photons. The particle nature-based indeterminacy on quantum feature generations observed in all HOM experiments is now removed by the intrinsic properties of paired photons' characteristics of symmetric detuning and detuning swapping between them. Even if the SPDC-generated photon pairs were replaced by independently coherent photon pairs, the same HOM dip-based anticorrelation could be achieved by coherence modification using an etalon set due to cavity optics. These coherence analyses for a quantum feature of a HOM dip are unprecedented and powerful enough to open the door to deterministic quantum information processing based on entangled photon pairs.

Although the photon detection randomness in each output port of a BS was given by the intrinsic property of the SPDC-generated photon pairs, it also could be classically manipulated using pure coherence optics in a coupled MZI scheme for macroscopic entanglement generation as recently suggested [40]. Compared with Fig. 1(b) of two incoherently coherent input photons, an MZI with a single input laser intrinsically results in perfectly coherent photon pairs with the $\pi/2$ phase difference, where entanglement can be later achieved via superposition between the split photons by the BS following in a cascaded MZI system [41]. Thus, the present investigation of photon characteristics for anticorrelation using coherence optics can be used for deterministic quantum manipulation without violating quantum mechanics. The equality between exclusive natures of a photon has already been predicted and observed in the Born rule tests conducted over the last decade [25-28]. Thus, a specific photon detection in the particle nature of a photon is now replaced by intensity correlation between ensemble coherence, where delay time is also replaced by a difference in phase.

**Conclusion**

We analyzed and numerically demonstrated the most important quantum feature of a Hong-Ou-Mandel dip for anticorrelation between paired entangled photons on a BS. In addition, a classical case was compared for a general scheme of bipartite quantum correlation, whose input fields were independently coherent. As a result, the fundamental physics of phase relationship between the paired photons was discovered. In fact, the uniform intensity output fields from the BS via two-photon correlation were not a necessary condition for the anticorrelation, but were for $g^{(2)}$ intensity correlation. In the classical case of independently coherent photons, deterministic control of the same anticorrelation was also analyzed and achieved if the coherence phase control between the paired photons were possible via an etalon set due to cavity optics. As demonstrated in the Born rule tests, the equivalence between the particle and wave natures of a photon supports the present analyses and results. In conclusion, the on-demand phase control of the quantum feature of anticorrelation has now been made available because the relative phase difference between the paired coherent photons can be controlled. Here, the control of relative phase difference between two paired photons is purely classical and does not violate quantum mechanics. For SPDC generated entangled photon pairs, such a deterministic control is not possible simply due to the intrinsic properties of the photon characteristics based on symmetric detuning between paired photons. The detuning swapping between paired entangled photons is an inherent property of SPDC processes different from the classical case of the coherent photons. Thus, the on-demand quantum correlation control using coherent photons has potential for deterministic quantum information processing.

Acknowledgments

This work was supported by GIST via GRI 2021.

Reference

1. J. Bell, "On the Einstein Poldosky Rosen paradox," *Phys.* **1**, 195 (1964).
2. C. K. Hong, Z. Y. Ou, and L. Mandel, "Measurement of subpicosocnd time intervals between two photons by interference," *Phys. Rev. Lett.* **59**, 2044 (1987).
3. A. Einstein, B. Podolsky, and N. Rosen, "Can quantum-mechanical description of physical reality be considered complete?" *Phys. Rev.* **47**, 777 (1935).
4. J. F. Clauser, M. A. Horne, A. Shimony, and R. A. Holt, "Proposed experiment to test local hidden-variable theories," *Phys. Rev. Lett.* **23**, 880 (1982).
5. R. Horodecki, P. Horodecki, M. Horodecki, and K. Horodecki, "Quantum entanglement," *Rev. Mod. Phys.* **81**, 865 (2009).
6. E. Knill, R. Laflamme, and G. J. Milburn, "A scheme for efficient quantum computation with linear optics," *Nature* **409**, 46 (2001).
7. P. Kok, W. J. Munro, K. Nemoto, T. C. Ralph, J. P. Dowling, and G. J. Milburn, "Linear optical quantum computing with photonic qubits," *Rev. Mod. Phys.* **79**, 135 (2007).
8. F. Arute, *et al.*, "Quantum supremacy using a programmable superconducting processor," *Nature* **574**, 505 (2019).
9. D. P. DiVinicenzo, "Quantum computation," *Science* **270**, 255 (1995).
10. L.-M. Duan and H. J. Kimble, "Scalable photonic quantum computation through cavity-assisted interactions," *Phys. Rev. Lett.* **92**, 127902 (2004).
11. N. Sangouard, C. Simon, H. de Riedmatten, and N. Gisin, "Quantum repeaters based on atomic ensembles and linear optics," *Rev. Mod. Phys.* **83**, 33 (2011).
12. N. Gisin, G. Ribordy, W. Tittel, and H. Zbinden, "Quantum cryptography," *Rev. Mod. Phys.* **74**, 145 (2002).
13. V. Scarani, H. Bechmann-Pasquinucci, N. J. Cerf, M. Dušek, N. Lütkenhaus, and M. Peev, "The security of practical quantum key distribution," *Rev. Mod. Phys.* **81**, 1301 (2009).
14. F. Xu, X. Ma, Q. Zhang, H.-K. Lo, and J.W. Pan, "Secure quantum key distribution with realistic devices," *Rev. Mod. Phys.* **92**, 025002 (2020).
15. N. T. Islam, C. C.W. Lim, C. Cahall, J. Kim, and D. J. Gauthier, "Provably secure and high-rate quantum key distribution with time-bin qudits," *Sci. Adv.* **3**, e17011491 (2017).
16. C. L. Degen, F. Reinhard, and P. Cappellaro, "Quantum sensing," *Rev. Mod. Phys.* **89**, 035002 (2017).
17. J. Jacobson, G. Gjork, I. Chung, and Y. Yamamoto, "Photonic de Broglie waves," *Phys. Rev. Lett.* **74**, 4835 (1995).
18. K. Edamatsu, R. Shimizu, and T. Itoh, "Measurement of the photonic de Broglie wavelength of entangled photon pairs generated by parametric down-conversion," *Phys. Rev. Lett.* **89**, 213601 (2002).
19. X.-L. Wang, *et al.* "18-qubit entanglement with six photons' three degree of freedom," *Phys. Rev. Lett.* **120**, 260502 (2018).

20. G. B. Lemos, V. G. Borish, D. Cole, S. Ramelow, R. Lapkiewicz, and A. Zeilinger, "Quantum imaging with undetected photons," *Nature* **512**, 409 (2014).
21. Y.-H. Deng *et al.* "Quantum interference between light sources separated by 150 million kilometers," *Phys. Rev. Lett*. **123**, 080401 (2019).
22. R. Lettow, Y. L. A. Reau, A. Renn, G. Zumofen, E. Ilonen, S. Götzinger, and V. Sandoghdar, "Quantum interference of tunably indistinguishable photons from remote organic molecules," *Phys. Rev. Lett*. **104**, 123605 (2010).
23. L. Mandel and E. Wolf, *Optical coherence and quantum optics, Ch. 8* (NY, Cambridge 1995).
24. R. D. Sorkin, "Quantum mechanics as quantum measure theory," *Mod. Phys. Lett.* **9**, 3119-3127 (1994).
25. U. Sinha, C. Couteau, T. Jennewein, R. Laflamme, and G. Weihs, "Ruling out multi-order interference in quantum mechanics," *Science* **329**, 418-420 (2010).
26. Bo-Sture K. Skagerstam, "On the three-slit experiment and quantum mechanics," *J. Phys. Commun*. **2**, 125014 (2018).
27. O. S. Magaña-Loaiza et al., "Exotic looped trajectories of photons in three-slit interference," *Nat. Communi*. **7**, 13987 (2016).
28. M.-O. Pleinert, A. Rueda, E. Lutz, and J. von Zanthier, "Testing higher order quantum interference with many-particle states," *Phys. Rev. Lett*. **126**, 190401 (2020).
29. P. G. Kwiat, K. Mattle, H. Weinfurter, and A. Zeilinger, "New high-intensity source of polarization-entangled photon pairs," *Phys. Rev. Lett.* **75**, 4337-4341 (1995).
30. M. Pant et al., "Routing entanglement in the quantum internet," *npj quantum info*. **5**, 25 (2019).
31. M. Pompili et al., "Realization of a multinode quantum network of remote solid-state qubits," *Science* **372**, 259-264 (2021).
32. B. S. Ham, "The origin of anticorrelation for photon bunching on a beam splitter," *Sci. Rep.* **10**, 7309 (2020).
33. E. Solano, R. L. de Matos Filho, and N. Zagury, "Deterministic Bell states and measurement of motional state of two trapped ions," *Phys. Rev. A* **59**, R2539 (1999).
34. P. G. Kwiat, A. M. Steinberg, and R. Y. Chiao, "High-visibility interference in a Bell-inequality experiment for energy and time," *Phys. Rev. A* **47**, R2472–R2475 (1993)
35. S. Kim and B. S. Ham, "Revisiting self-interference in Young's double-slit experiments," arXiv:2104.08007 (2021).
36. C. Couteau, "Spontaneous parameteric down-conversion," *Cont. Phys.* **59**, 291-304 (2018).
37. S. Volkovich and S. Shwartz, "Subattosend x-ray Hong-Ou-Mandel metrology," *Opt. Lett.* **45**, 2728-2731 (2020).
38. F. Dieleman, M. S. Tame, Y. Sonnefraud, M. S. Kim, and S. A. Maier, "Experimental verification of entanglement generated in a plasmonic system," *Nano Lett.* **17**, 7455-7461 (2017).
39. F. Bouchard et al., "Two-photon interference: the Hong-Ou-Mande effect," *Rep. Prog. Phys.* **84**, 012402 (2020).
40. B. S. Ham, "Macroscopically entangled light fields," *Sci. Rep.* (To be published).
41. V. Degiorgio, "Phase shift between the transmitted and the reflected optical fields of a semireflecting lossless mirror is π/2," *Am. J. Phys.* **48**, 81–82 (1980).